\documentclass[12pt]{article}
\usepackage{natbib}
\usepackage{amsmath}
\usepackage{graphicx}
\usepackage[inline]{enumitem}
\usepackage{enumerate}
\usepackage{url} 
\usepackage{mathtools}

\usepackage{amsthm}
\usepackage[dvipsnames]{xcolor}

\usepackage{float}
\usepackage{enumitem}
\usepackage{amsfonts}
\usepackage[ruled,vlined]{algorithm2e}
\usepackage{algpseudocode}
\usepackage{multirow}
\usepackage{ulem}
\newcommand{\blind}{1}

\begin{document}

\def\spacingset#1{\renewcommand{\baselinestretch}%
{#1}\small\normalsize} \spacingset{1}


\if1\blind
{
  \title{\bf Bayesian Nonparametric Multivariate Mixture of Autoregressive Processes: With Application to Brain Signals}
  \author{Guillermo C. Granados-Garcia\\
    King Abdullah University of Science and Technology (KAUST) \\
    Raquel Prado \\
    Univresity of California, Santa Cruz \\
        Hernando Ombao \\
    King Abdullah University of Science and Technology (KAUST) }
  \maketitle
} \fi

\if0\blind
{
  \bigskip
  \bigskip
  \bigskip
  \begin{center}
    {\LARGE\bf Novel decomposition of multivariate signals in its fundamental waveforms via Bayesian non-parametric mixtures of autoregressive kernels}
\end{center}
  \medskip
} \fi

\begin{abstract}
One of the goals of neuroscience is to study interactions between different brain regions during rest and while performing specific cognitive tasks. The Multivariate Bayesian Autoregressive Decomposition (MBMARD) is proposed as an intuitive and novel Bayesian non-parametric model to represent high-dimensional signals as a low-dimensional mixture of univariate uncorrelated latent oscillations. Each latent oscillation captures a specific underlying oscillatory activity and hence will be modeled as a unique second-order autoregressive process due to a compelling property that its spectral density has a shape characterized by a unique frequency peak and bandwidth, which are parameterized by a location and a scale parameter. The posterior distributions of the parameters of the latent oscillations are computed via a  metropolis-within-Gibbs algorithm. One of the advantages of MBMARD is its robustness against misspecification of standard models which is demonstrated in simulation studies. The main scientific questions addressed by MBMARD are the effects of long-term abuse of alcohol consumption on memory by analyzing EEG records of alcoholic and non-alcoholic subjects performing a visual recognition experiment. The MBMARD model exhibited novel interesting findings including identifying subject-specific clusters of low and high-frequency oscillations among different brain regions. 
\end{abstract}

\noindent%
\spacingset{1.5} 

\section{Introduction}

In the last decades, neuroscience has significantly advanced our understanding of the brain’s encoding and retrieving processes, aided by novel statistical models and data-analytic techniques. Modern neuroscience utilizes electroencephalogram (EEG) records as a non-invasive technique to study the dynamics and associations of the oscillatory patterns in brain electrical activity during resting state and various cognitive activities. Thus, the EEG signals recorded across the scalp can be interpreted as a mixture of latent sources oscillating at various frequencies with random amplitudes.

The basic oscillatory patterns in a stationary process can be formulated as having a spectral density characterized by different peaks determined by the location of the frequency peaks and sharpness (or bandwidth) around these peaks.These features can serve as biomarkers to determine differences among populations; for instance, \cite{dickinson2018peak} analyzed the maximal peak location as a biomarker of cognitive function across the autism spectrum. 

The signal waves are usually decomposed into separate frequency bands obtained empirically based on practical considerations \citep{ buzaki}. Studies have demonstrated the association between sensory and cognitive processing and the power over frequency bands. Examples of these studies include memory \citep{klimesch1997eeg}, language translation \citep{grabner2007event} or intelligence \citep{jauvsovec2000differences}. However, individual differences across subjects and other factors influence the specific frequency of task-associated oscillations, making the assumption of a universal frequency band partition unrealistic. This variation of the spectral pattern across the population on their peaks location and spread of power motivated the need to develop novel methods to determine frequency bands (\cite{doppelmayr1998individual}, \cite{brucekrafty}). 

The standard spectral approach to analyzing multivariate time series starts by transforming the entire multi-channel recordings into the spectral domain through the discrete Fourier transform (\cite{shumway2017time}, \cite{ombao2021spectral}). This approach identifies the most prominent oscillations or waves that contribute the most to the signal variance. Indeed, the cross-association information between different signals in the frequency domain is contained in the spectral matrix. Different quantities derived from the spectral matrix can measure the strength of cross-channel relations as the coherence, coherency, partial directed coherence (PDC) \cite{baccala_partial_2001}, among others, see \cite{ombao2021spectral} and \cite{ombaosebastien2008} for a comprehensive review of spectral dependence metrics. 

There are many approaches to estimating spectral matrix including tapering and smoothing of the periodogram matrices,  \citep{shumway2017time, Brockwell:1986:TST:17326}, smoothing the cross-correlation matrix via a multivariate window lag functions (\cite{wei_multivariate_2019,shumway2017time}). Other approaches smooth the Cholesky decomposition by using splines \citep{dai_multivariate_2004}, Bayesian estimation \citep{rosen_automatic_2007}, penalization of the Whittle likelihood \citep{krafty_penalized_2013}, non-parametric Bayesian mixture based on Bernstein polynomials (\cite{petrone1999bayesian}, \cite{Choudhuri}, \cite {MEIER2020104560}, \cite{hart_nonparametric_2022}). These approaches are not ideal when the goal is to identify oscillatory sources as spectrum peaks since these methods focus on estimating the spectrum as a whole, often resulting in over-smoothing the peaks. 

Another approach to analyze EEG waveforms are factor models; this models describe a similar approach to our model by assuming there are latent processes that describe a complex time series. Factor models have their development in dynamic linear models (DLM) as described in \cite{west_bayesian_1995}, \cite{west1997time}, and \cite{motta2012evolutionary}, where a non-stationary time series is decomposed into latent processes. Each of these latent processes can be stationary, describing cyclical processes. Other components can be non-stationary with time-varying mixing matrices. 

In DLM, the latent processes are also known as factors that capture specific effects on brain electrical activity following cognitive stimulation or during resting state. DLM models were extended to time-varying vector autoregressive processes (TVAR) as latent factors to analyze EEG traces displaying non-stationary behavior for subjects under diverse experimental conditions (see \cite{prado1998latent}, \cite{west_evaluation_1999}, \cite{krystal1999new}, \cite{prado2001multichannel}, \cite{prado2002time}). Recently \cite{li_modeling_2020} investigate the oscillatory behavior of local field potentials of rats using a factor model with Gaussian processes as components.

The main advantage of factor models is their predictive power  and ease in interpretability via the decomposition of complex non-stationary behaviors in simpler processes. The main difference with our approach is our target problem in analyzing brain signals. Moreover, while factor models have been used to model time series of longer duration, the focus of the proposed model is on the immediate reaction of the brain during stimuli. The problem is that due to the speed in which the brain operates, capturing the effect of stimuli in the brain requires an immediate analysis after a stimulus.

The novel model proposed here to investigate the cross-dependence of oscillatory patterns in stationary time series is the multivariate Bayesian mixture autoregressive decomposition (MBMARD). Inspired by \cite{gao2016evolutionary} and building on \cite{granados2021brain}, the proposed MBMARD model captures the association across the channels based on a Dirichlet process mixture (DPM). The DMP characterizes the strength of the shared oscillatory waves as latent second-order autoregressive processes. 

The development of MBMARD targets is to investigate and compare the oscillations of the EEG signals from a long-term alcoholic subject and a non-alcoholic subject during a visual memory experiment. Evidence indicates that long-term alcoholics show deficits in response time and demand higher effort at retrieving visual information (see \cite{zhanga_is_1997, zhang_electrophysiological_1997}).

The \underline{novel} elements of the MBMARD are
a) The estimates allow the identification of specific oscillations that are shared across many channels. 
b) Each mixture component is associated with a unique wave in multivariate records.
c) Since the kernels are derived from a parametric times series model, the computation of statistical quantities in the frequency domain, time domain, and Granger causality measures is straightforward. The \underline{advantages} of MBMARD over existing approaches are
a) The estimator reduces the absolute integrated error compared to the Bernstein polynomial approach.
b) The model parameters are associated with biomarkers from the neuroscientist literature. 
c) MBMARD is computationally scalable since it focuses on identifying a reduced number of latent sources and not relying on a higher number of baseline functions to improve accuracy.

The structure of the remainder of this paper is as follows. The fundamentals of the MBMARD model as the mixture model and spectral basis developed from a second-order autoregressive process are introduced in Section 2. Section 3 presents important properties of MBMARD such as cross-correlation, spectral matrix, and causality measures. In Section 4, Bayesian inference and prior specification are developed. In Section 5, the performance of MBMARD is evaluated in a simulation study comparing to one competitor method for two scenarios. Section 6 uses the MBMARD method to analyze alcoholic and non-alcoholic subjects while performing a visual recognition experiment finding the frontal and occipital zones are associated with low-frequency oscillations while temporal zones are associated with high-frequency sources.

\section{Mixture model for multivariate time series}

The multivariate Bayesian mixture autoregressive decomposition (MBMARD) framework will be developed in this section. The proposed MBMARD method will be employed to identify EEG oscillatory behaviors at low and high frequencies often seen superposed at different frequencies and amplitudes. The EEGs will be represented as a mixture of latent quasi-periodic processes. These latent processes are modeled as low-order autoregressive processes (\cite{gao2016evolutionary}, \cite{granados2021brain}, \cite{west_evaluation_1999} ). 
 This approach is also justified by the fact that the spectral density of a stationary time series could be approximated by the spectral density of an autoregressive process of order $p$ (AR($p$)) where $p$ is sufficiently large, depending on the complexity of the spectra as characterized by number of peaks and the corresponding bandwidths (see \cite{gao2016evolutionary}, \cite{granados2021brain} and \cite{shumway2017time}). Second, a fundamental property of an AR($p$) process is that its spectrum can be decomposed as a superposition of the spectra from AR(1) and AR(2) processes. This univariate model is extended to the multivariate setting based upon idea that multivariate time series with correlated components have shared waveforms, which characterizes the synchronization of subsets of univariate signals.

\subsection{Autoregressive latent oscillations}

\cite{granados2021brain} demonstrated that the latent sources $Z_j(t), j=1,...,K$ of a signal can be modeled by uncorrelated causal AR(2) processes $Z_j(t)=\phi_{1,j}Z_j(t)+\phi_{2,j} Z_j(t-2)+\epsilon_j(t)$ with $\epsilon_j(t)$ i.i.d. white noise process. The reason for using an AR(2) is its spectral density is characterized by a single peak function with peak location $\psi_j \ \in (0,0.5)$ and bandwidth $M_j>1$. The derivation of the basis function for the mixture model as a proper probability density function (PDF) and its parameters is as follows. 

First the characteristic polynomial of the AR(2) $1-\phi_{1,j}\lambda-\phi_{2,j}\lambda^2$ 
 is assumed to have complex conjugate roots $\lambda^*=M_j \exp(\mp 2\pi i \psi_j)$. Based on the roots of the AR(2) characteristic polynomial the following one-to-one relationship holds: $\phi_{1,j}=\frac{2}{M_j}cos(2\pi \psi_j)$ and  $\phi_{2,j}=-\frac{1}{M_j^2}$.   

Finally, the basis function is obtained by dividing the spectral density of the AR(2) by its variance. Then the derived PDF associated with $Z_j$ is 
\[g_j(\omega; \psi_j,L_j)= \frac{2(1-e^{-2L_j})((1+e^{-2L_j})^2-4\cos^2(2\pi\psi_j)e^{-2L_j} )}{(1+e^{-2L_j})|1- 2\cos(2\pi \psi_j)e^{-L_j}(e^{-i2\pi\omega}) +e^{-2L}(e^{-i4\pi\omega})|^2}\]
Where $L_j=log(M_j)>0$, to set appropriate priors on the bandwidth parameters.
\begin{figure}[H]
    \centering
    \includegraphics[width=16cm, height=5cm]{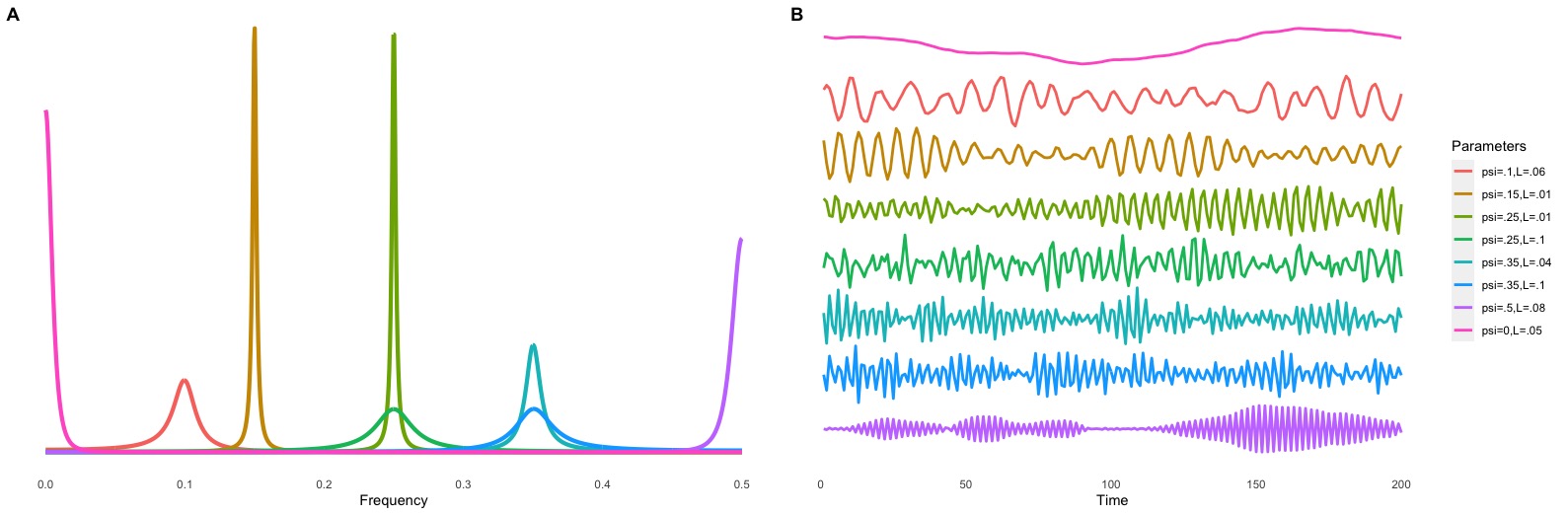}
    \caption{ \textbf{A:} Examples of the autoregressive kernel with different values for the phase $\psi$ (location) and bandwidth (scale) parameters $L$. \textbf{B:} Generated simulations from the AR(2) processes defined by each parameter combination.   }
    \label{fig:kernelexample}
\end{figure}

Figure~\ref{fig:kernelexample} demonstrate the flexibility of the AR(2) spectral density showing seven examples of the autoregressive spectral density under different combinations of $\psi$ and $M$. The phase parameter defines the location of the peak across the frequencies in the interval $(0,.5)$. The associated waveform observed on the right plot shows that the waves generated are from low-frequency to high-frequency oscillations.  The bandwidth parameter determines the sharpness of the kernel. Note that sharper peaks correspond to low values of $M$ and are associated with quasi-periodic behavior and that the range of the frequency axis can be transformed from $0$ to $0.5 \times$ the sampling rate. 

\subsection{Mixture model}
We now provide the general definition of the mixture that will be vital to the derivation of the statistical dependence measures. Let $X(t)=\{X_1(t),\ldots,X_n(t)\}$ be a multivariate time series which is assumed to be a mixture of latent oscillations $Z(t)=(Z_1(t), \ldots, Z_K(t))$, where the number of oscillations $K$ is unknown. Note that when target dimension reduction is biologically justifiable, a usual condition is $K <<< n$. However, univariate processes ($n=1$) can be composed of several latent waveforms ($K>1$) \citep{gao2016evolutionary, granados2021brain}. Moreover, as higher dimensions $n>>1$ of EEG channels are considered, the model will result in a parsimonious representation of few latent waveforms.

Based upon the assumption the set of latent processes are uncorrelated AR(2) processes; the observed multichannel time series is represented as a mixture $Z(t)=A_1 Z(t-1)+A_2 Z(t-1)+\epsilon(t)$ where $A_1=diag(\phi_{1,1},\cdots,\phi_{1,K})$, $A_2=diag(\phi_{2,1},\cdots,\phi_{2,K})$, and  $\epsilon(t)$ is a white noise process with a diagonal autocovariance matrix $\Sigma=diag(\sigma_{1}^2,\cdots,\sigma_{p}^2)$. The general representation of the described linear mixture model is $X(t)=\Lambda Z(t)+q(t)$. Where $q(t)$ is a zero-mean white noise process of dimension $n$ with covariance matrix $\Sigma_q=\sigma^2_q I_{n \times n}$ where $I_{n \times n}$ is the identity matrix of $n \times n$ and $\Lambda$ is a $n \times K$ matrix containing the mixture weights such that $\sum_{j=1}^K \lambda_{ij}^2=1, \ i=\{1,\ldots,n\} $ for identifiability. Thus, the weights $\Lambda=\{\lambda_{i,j}\}$ $0 \leq\lambda_{i,j}\leq1$ represent the square root of the contribution of the $j$-th latent oscillation to the total variance of the $i$th series. 

\subsection{MBMARD model autocorrelation}
One advantage of the MBMARD approach is the closed form derivation of statistical quantities of the multivariate signal in terms of the latent processes parameters ${\psi_j, M_j}, j=1,\ldots, K$ leading to a faster computation.
The autocovariance $\Sigma_{XX}(h)$ of $X(t)$ can be expressed in terms of the spectral matrix and autocovariance $\Sigma_{ZZ}(h)$ of $Z(t)$ as follows:
$
\Sigma_{XX}(h)=\Lambda \Sigma_{ZZ}(h)\Lambda^T +\sigma^2_q I_{n \times n}
$, where $\Sigma_{ZZ}(h)=diag(\gamma_1(h),\ldots,\gamma_K(h))$ and $\gamma_{Z_j}(h)=Cov(Z_j(t),Z_j(t+h))$ is the autocovariance function of the $j$-th latent component. 

Note this implies the cross-correlation between the univariate signal components $X_m(t)$, $X_\ell(t)$ is $\gamma_{m\ell}(h)=\sum_{j=1}^K \lambda_{mj}\lambda_{\ell j}\gamma_j(h)$. Thus, the association between channels is determined by the mixture of the covariance function of each latent source wave. The cross-covariance of the latent components with the signals is $Cov(X_i(t),Z_j(t+h))=\lambda_{ij}\gamma_j(h)$. The covariance matrix  $\Sigma_{Z,X}(h)$ that includes the latent components and the signals, can be expressed as the following block matrix  
\[
\Sigma_{Z,X}(h) 
=
\begin{pmatrix}
\Sigma_{ZZ}(h) &  \Sigma_{ZZ}(h) \Lambda^T \\
\Lambda \Sigma_{ZZ}(h) & \Lambda \Sigma_{ZZ}(h)\Lambda^T+\sigma^2_q I_{n \times n}.
\end{pmatrix}
\]

\subsection{MBMARD spectral matrix computation}

Analogous to the time domain approach, the association between the univariate signals in the frequency domain is contained in the spectral matrix $S_{XX}(\omega)$ computed from the spectral matrix of the latent processes $S_{ZZ}(\omega)=diag(g_{1}(\omega),\ldots,g_{K}(\omega) )$, with $g_{j}(\omega)$ $j=1,\ldots,K$ computed as the autoregressive kernels which is diagonal by the assumption of uncorrelated error terms of $Z(t)$. 
$
S_{XX}(\omega)= \Lambda S_{ZZ}(\omega) \Lambda^T   +\sigma^2_q I_{n \times n}  
$
The cross-spectrum between time series $X_m(t)$, $X_\ell(t)$ is $S_{X_{m},X_{\ell}}(\omega)=\sum_{j=1}^K \lambda_{mj}\lambda_{\ell j}g_j(\omega)+\sigma^2_q$. The cross-spectrum between the signal $X_i(t)$ and  the latent process $Z_j(t)$ is $S_{X_{i},Z_{j}}(\omega)=\lambda_{ij}g_j(\omega)$. Analogous to the time domain computations the spectral matrix of the latent processes and the multivariate signal is represented by the block matrix $S_{Z,X}(\omega)$.
\[
S_{Z,X}(\omega) 
=
\begin{pmatrix}
S_{ZZ}(\omega) &  S_{ZZ}(h) \Lambda^T \\
\Lambda S_{ZZ}(h) & \Lambda S_{ZZ}(h)\Lambda^T+\sigma^2_q I_{n \times n}
\end{pmatrix}
\]

One specific measure of dependence between channels $X_m(t)$, $X_l(t)$ is coherence (at a particular frequency $\omega$) which is given by
$\frac{ |S_{X_{m},X_{\ell}}(\omega)|^2 }{S_{X_{m},X_{m}}(\omega)S_{X_{\ell},X_{\ell}}(\omega)}$ which takes values closer to 1 as $\frac{ \sigma^2_q}{|\Lambda S_{ZZ}(\omega) \Lambda^T   +\sigma^2_q I_{n \times n} |} \rightarrow 0$ and in terms of each associated latent oscillation the contribution to the coherence is considerable if their weights are similar, otherwise the cross-spectrum will be shrunk by low weights.

\subsection{Granger causality analysis via MBMARD}

Beyond quantities such as cross-correlation and coherence, recent literature also aims to measure more general lead-lag relationships between brain signals through Granger causality (\cite{kaminski_new_1991},  \cite{baccala_partial_2001},\cite{liu_statistical_2021}, \cite{ombao2021spectral}, \cite{pinto2021scau} ).

 The concept of causality has been debated from a philosophical point of view to its practical use and interpretation in the statistical world ( \cite{baldi_bayesian_2020}). According to \cite{russell_inotion_1913} a way to understand causality in a probabilistic context is if a sequence of events is observed frequently, then there is a probability that a cause-and-effect relationship exists. Recently, \cite{baldi_bayesian_2020} proposed a framework that treats the causality as a hypothesis that could be tested through statistical models. Causality is inferred by considering two events, A as a possible cause and B as a potential effect. Under this framework, causality is stated as follows: given event B, what is the probability that event A is the cause; or given the occurrence of A, what is the probability that event B follows. 

Causality in time series analysis was first reported in  \cite{granger_economic_1963} from the approach of minimizing prediction uncertainty measured by the variance. To formalize,  let $X(t) =X_1(t),\ldots, X_n (t)$ be a multivariate time series and denote $Q$ to be the all series of $X$ and $Q_k$ the set that excludes $X_k$. Suppose that our goal is to predict a series $X_i(t)$. Suppose now that the optimal unbiased estimator $P(Q)$, based on the whole set $Q$, exists and has variance $V(Q)$ and $P(Q_k)$ is the optimal predictor based on $Q$ with variance $V(Q_k)$. Then, a measure of the extent of causality from $X_k(t)$ to $X_i(t)$ denoted as $X_k \rightarrow X_i$ is the relative reduction in the variance between $V(Q_k)$ and $V(Q)$.

Further work by \cite{geweke_measurement_1982}, \cite{hosoya_decomposition_1991} and \cite{ombao2021spectral} have imported Granger causality concepts into the frequency domain.  Frequency domain causality is of particular interest when the goal is to identify lead-lag dependence of the quasi-cyclical behaviors between the signals.


The seminal work of \cite{kaminski_new_1991} investigated Granger causality on brain signals via the directed transfer function (DTF) by assuming an autoregressive model for a multivariate time series $X(t)=(X_1(t),\cdots,X_n(t)$ as $\sum_{r=0}^R A_r x(t-r) $ where $A_r$ is the matrix of coefficients at lag $r$. 
Moreover, in \cite{baccala_partial_2001}, the DTF is interpreted as the quantity that measures the information flow between two signals, including indirect influences from the rest of the multivariate set. Consequently,  partial directed coherence (PDC) is defined to be $\pi(\omega)=\{\pi_{ij}\}$ as an alternative function to measure the direct influence $X_j \rightarrow X_i$ filtering other sources information flows.

In \cite{liu_statistical_2021}, an approach is developed for investigating the temporal evolution of Granger causality and directionality (information transfer) from one time series to another. The local measure is analogous to Hosoya's metric computed across time, assuming that the variation of the multivariate time series oscillatory behavior changes slowly. Granger Causality measures have been applied in fMRI data to explain the effective connectivity of specific regions of interest over other target voxels without assuming a structural model a priori (see \cite{lindquist_statistical_2008}, \cite{deshpande_multivariate_2009}, \cite{roebroeck_mapping_2005} ).  
\begin{figure}[H]
    \centering
    \includegraphics[width=12cm]{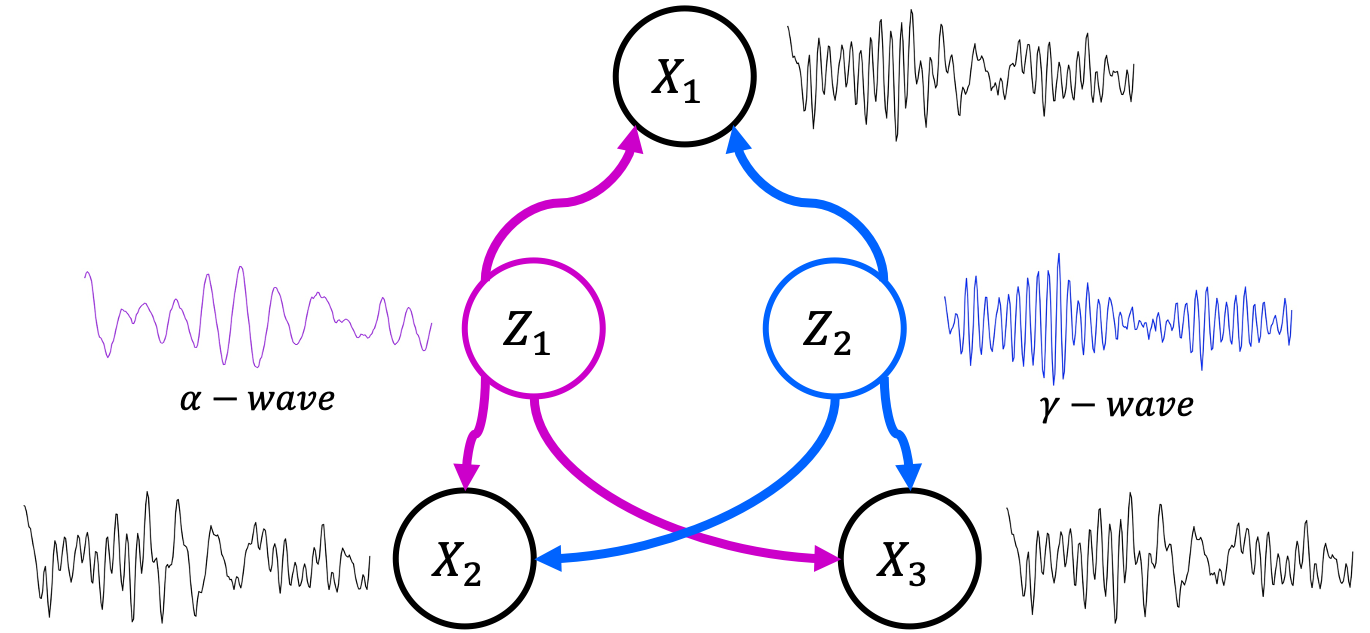}
    \caption{ Causality model: The observed signals $X(t)$ are assumed as a linear combination of the latent waveforms $Z(t)$. The relation among the signals $X1(t), X_2(t), X_3(t)$ is indirect via the latent oscillations. The model explains the synchronization of the signals. The signals are related indirectly via the latent oscillations.}
    \label{fig:causalitysketch}
\end{figure}

In the MBMARD framework, Granger causality from a latent oscillation $Z_j(t)$ to an observed signal $X_i(t)$ is computed as 
\[|\pi_{ij}(\omega)|= \frac{\lambda_{ij} |\Phi_j(\omega)| }{\sqrt{|\Phi_j(\omega) |^2 \sum_{i=1}^n \lambda^2_{ij} +|1-\Phi_j(\omega)|^2  } }\] 
where $\Phi_j(\omega)=\phi_{j1}e^{-i2\pi\omega}+ \phi_{j2}e^{-i4\pi\omega}$. Figure \ref{fig:causalitysketch}  provides a sketch of the MBMARD framework with two latent waves oscillating at different frequencies and their mixture providing a realistic approximate behavior of the observed signals. 

In the the model the indirect association between signals is captured by the latent oscillations explaining the apparent synchronization of the signals. Thus, the PDC between two distinct signals will be zero, because the ``partial" definition of the PDC removes the direct influence of the latent waveforms. 

\section{Inference under the non-parametric Bayesian model}

Let $\{X_i(t)\}_{i=1}^n, t=1,\ldots,T$, be an observed a multivariate time series of length $T$ where is even, for simplicity and $n$ denotes the number of channels. The first step is to standardize each of the series to have mean 0 and variance 1. The discrete Fourier transform (DFT) of the data is computed as $d(\omega_m)=T^{-1/2} \sum_{t=1}^T X(t)\exp(-2\pi i \omega_m t)$, $ (m=1,\ldots, \frac{T}{2} -1)$ where $\omega_m=m/T$ and $d(\omega_m)$ is a vector of Fourier coefficients with dimension $n$ (one Fourier coefficient at frequency $\omega_m$ for each channel). The approximate log-likelihood up to a constant is the multivariate Whittle likelihood (\cite{whittle_analysis_1953}) given by 
\[
\ell(S_{XX})=-\sum_m^{(T-1)/2}\left\{ log|S_{XX}(\omega_m)| + d(\omega_m)^* S_{XX}^{-1}(\omega_m)d(\omega_m) \right\}
\]
This log-likelihood assumes Under weak stationarity the $d(\omega_m)$ are asymptotically uncorrelated across Fourier frequencies distributed as complex Normal distributed with mean 0 and variance covariance matrix $S_{XX}(\omega_m)$ which is Hermitian and complex-valued.

The matrix $S_{XX}(\omega_m)$ is modeled through the Bayesian non-parametric model by assuming the spectral matrix as an infinite dimensional parameter coming from the mixture model parameterized through the vector $\theta=\{ \Lambda, \bar{\psi}, \bar{L} , K, \sigma^2_q  \} $.  $\theta$ is assumed distributed from a measure $G$ and set to it a Dirichlet Process prior (DP) with precision parameter $\alpha$ (\cite{ferguson1973bayesian}). 
\begin{gather*}\label{eqdpmodel}
d(\omega_m) \ \sim -log|S_{XX}(\omega_m)| - d(\omega_m)^* S_{XX}^{-1}(\omega_m)d(\omega_m)   , \\
S_{XX}|\theta = \Lambda S_{ZZ}(\omega) \Lambda^T   +\sigma^2_q I_{n \times n} , \quad
\theta |G \sim G, \quad
G \sim DP(G_0,\alpha).
\end{gather*}
The necessary constraints to achieve identifiability are discussed below along with the prior specifications and the inference algorithm. 

\subsection{Prior specifications}
The proposed algorithm adapts the matrix stick breaking representation from \cite{dunson_matrix_2008} to compute the posterior distributions of the mixing weights matrix $\Lambda$. The iterative algorithm is based on the update of two matrices $V_{n \times v}$ and $\Theta_{n \times v}$ where $v$ is the truncation level of the DP introduced by \cite{ishwaran2002approximate} as a finite approximation of the constructive stick-breacking representation of \cite{sethuraman1994constructive}.

Each row of the matrices represents the univariate form of the DP, i.e., $\Theta_{j,1:v}$ represents a set of atoms associated with the channel $j$ and $V_{j,1:v}$ represents the individual weights of the atoms, such that, given a value for the number of components $K^*$ the $k$-th weight in channel $j$ is computed as $p_{jk}=\sum_{h=1}^v V_{j,h}I( \frac{k-1}{K^*} < \Theta_{j,h} \leq \frac{k}{K^*})$. The phase parameters $\psi_j$ are constrained by a random partition built from an auxiliary variable vector $B = [b_0, \ldots, b_K]$, where $b_0 < b_1 < \ldots, b_K$, $b_0=0$, $b_K = 0.5$ such that $b_{j-1}<\psi_j<b_j$. 

The MH within Gibbs algorithm, implemented for posterior sampling, along $S$ iterations performs a Birth-Death process. At the $s$-th iteration a subinterval $\mathcal{I}_{j^*}$ of the partition $B$ is selected at random with equal probability. Then the algorithm decides with $0.5$ probability if the birth of a new component should be done by splitting $\mathcal{I}_{j^*}$ adding a new value within the subinterval. Otherwise the death process joins $\mathcal{I}_{j^*}$ with a neighbor subinterval deleting any of the edges of the subinterval, and generating new parameters for the resulting component. This process updates the number of components $K$ at each iteration allowing to track the uncertainty on $K$ and ensuring the identification of a peak per latent oscillation.

A uniform prior is set to $\psi_j$ inside its corresponding subinterval. The partition cutoffs are set a uniform prior between its neighbor values. For the log-modulus parameters $log(M_j)>=0$ the Jeffrey's prior $log^{-2}(M_j)$ is chosen. The prior for the DP scale parameter $\alpha$ was set to gamma distribution $\Gamma(a,b)$ with $a=.1$, $b=.1$ to set it a weakly informative prior.

In this case the marginal posterior of $\alpha$ was implemented based on the derivations in \cite{escobar1995bayesian}. The update of the parameters is made individually with symmetric proposals using Metropolis-Hastings steps. 

In applied Bayesian analysis, prior distribution choice can impact the estimations of the parameters (\cite{depaoli_importance_2020}). Prior sensitivity analyses were performed for the simulation study and the EEG data analysis under the following alternative priors for model parameters. 

The location parameters $\psi_j$ alternative prior is a beta distribution of four parameters with shape parameters set equal to 2 to keep the location parameters constrained to each subsegment and centered a priori, i.e., $\psi_{j}\sim Beta(2,2,\epsilon_{j-1}, \epsilon_{j})$. The Bandwidth parameters $L_j$ were set to a uniform prior  $L_j\sim U(0,2)$ as the alternative, where the upper limit serves to cap broad kernels. The alternative prior to the number of components is a Uniform discrete constrained to a maximum number of components $K_{max}$, i.e., $K\sim U(0, K_{max})$ where $K_{max}$ was set equal to 30. The results of the simulations and data analysis (not shown here) indicate that the prior choice does not significantly impact the final results since, in all combinations of priors, the Gibbs within MH algorithm converges to the same estimates after the same number of iterations.  

\section{Simulation Study}
 
The proposed MBMARD method and its non-parametric Bayesian inference algorithm is tested in four different simulation settings. The first setting is a sanity check to illustrate the MCMC where four different AR(2) $Z(t)={Z_1(t), Z_2(t), Z_3(t), Z_4(t)}$ processes are generated with phase parameters $\psi_1=.005$, $\psi_2=.03$, $\psi_3=.06$, $\psi_4=.3$ and log modulus parameters $log(M_j)=0.03 \ j={1,2,3,4}$. Based on the true latent processes a multivariate set of 7 time series is generated as follows 
\begin{equation}
\begin{pmatrix}
 X_1(t) \\
 X_2(t) \\
 X_3(t) \\
 X_4(t) \\
 X_5(t) \\
 X_6(t) \\
 X_7(t)
\end{pmatrix}=
\begin{pmatrix}
1 & 0 & 0 & 0 \\ 
0 & 1 & 0 & 0 \\ 
0 & 0 & 1 & 0 \\ 
0 & 0 & 0 & 1 \\ 
0.4^{1/2} & 0 & 0.6^{1/2} & 0 \\ 
0 & 0.7^{1/2} & 0 & 0.3^{1/2} \\ 
0 & 0 & 0.3^{1/2} & 0.7^{1/2}
\end{pmatrix}
\begin{pmatrix}
Z_1(t)  \\
Z_2(t)  \\
Z_3(t)  \\
Z_4(t)  \\
\end{pmatrix}+
.1 \epsilon
\label{simuscen1ar12}
\end{equation}
Where $\epsilon$ is a set of 7, i.i.d standard Gaussian noise variables, this example represents processes with peaks concentrated at low frequencies, which is what is most found in the data sets considered for the applications.

The second simulation setting, consider as latent oscillations a high-order autoregressive process, $Z_1(t)=0.9Z_1(t-4) + 0.7Z_1(t-8) -0.63Z_1(t-12) + \epsilon_1(t)$, a moving average process $Z_2(t)=-.3\epsilon_2(t-4)-.6\epsilon_2(t-3)-.3\epsilon_2(t-2)+.6\epsilon_2(t-1)+\epsilon_1(t)$, and two low-order autoregressive processes, one concentrated in low frequencies  $Z_3(t)=.8Z_3(t-1)+\epsilon_3(t)$ and other generated to display high frequencies $Z_4(t)=-.8Z_4(t-1)+\epsilon_4(t)$; along with 3 time series that mix the latent oscillations. The generated time series is as follows: 
\begin{equation}
\begin{pmatrix}
 X_1(t) \\
 X_2(t) \\
 X_3(t) \\
 X_4(t) \\
 X_5(t) \\
 X_6(t) \\
 X_7(t)
\end{pmatrix}=
\begin{pmatrix}
1 & 0 & 0 & 0 \\ 
0 & 1 & 0 & 0 \\ 
0 & 0 & 1 & 0 \\ 
0 & 0 & 0 & 1 \\ 
0 & 0 & 0.4^{1/2} & 0.6^{1/2} \\ 
0 & 0.6^{1/2} & 0.4^{1/2} & 0 \\ 
0 & 0.5^{1/2} & 0 & 0.5^{1/2}

\end{pmatrix}
\begin{pmatrix}
Z_1(t)  \\
Z_2(t)  \\
Z_3(t)  \\
Z_4(t)  \\
\end{pmatrix}+
.1 \epsilon
\label{simuscen1mix}
\end{equation}

The proposed method is compared with the method in \cite{MEIER2020104560} that uses Bernstein polynomial splines extended to matrices and propose to model the weight matrix induced by a Hermitian positive definite Gamma process. The computations over the simulated scenarios were made through the R package \textbf{ beyondWhittle} ( \cite{beyondWhittle}).  

The Metropolis-Hastings within Gibbs algorithm was implemented in Rcpp \citep{rcpp1} to boost the computational performance. The MBMARD method was run for each simulation setting for 50000 total iterations and three chains. The process to summarize all the chains is the following: first, a burn-in period is defined. For the after burn-in samples kept, the likelihood values of all chains are stacked together to identify the samples whose log-likelihood is higher than the $95\%$ quantile of the stacked values.  A matrix with $N+2$ columns is created, where the first two columns correspond to the stacked values vertically of the phase parameter and the bandwidth parameter, respectively. The other $N$ columns are filled similarly with the estimated weights, for all the iterations and chains found in the last step. 

The matrix of stacked values generates a cloud of points in a $N+2$ dimensional space. The next step is to identify the components frequently found by the MCMC algorithm. To this end, a clustering method from the R package  \textbf{mclust} (\cite{mclust}) is run considering all possible number of components of each iteration then the optimized configuration for the complete integrated likelihood is chosen.

The algorithm returns, for each cluster, the mean and standard deviation of each of the parameters associated with the latent oscillatory waves. Finally, to compute the estimated spectral matrix, the iterations corresponding to the highest likelihood values obtained along the algorithm runs denoted as  are set together and used to compute the pointwise mean for all the elements of the spectral matrix at all frequencies.
\subsection{Simulation results}

\begin{figure}[H]
    \includegraphics[width=16cm,height=6cm]{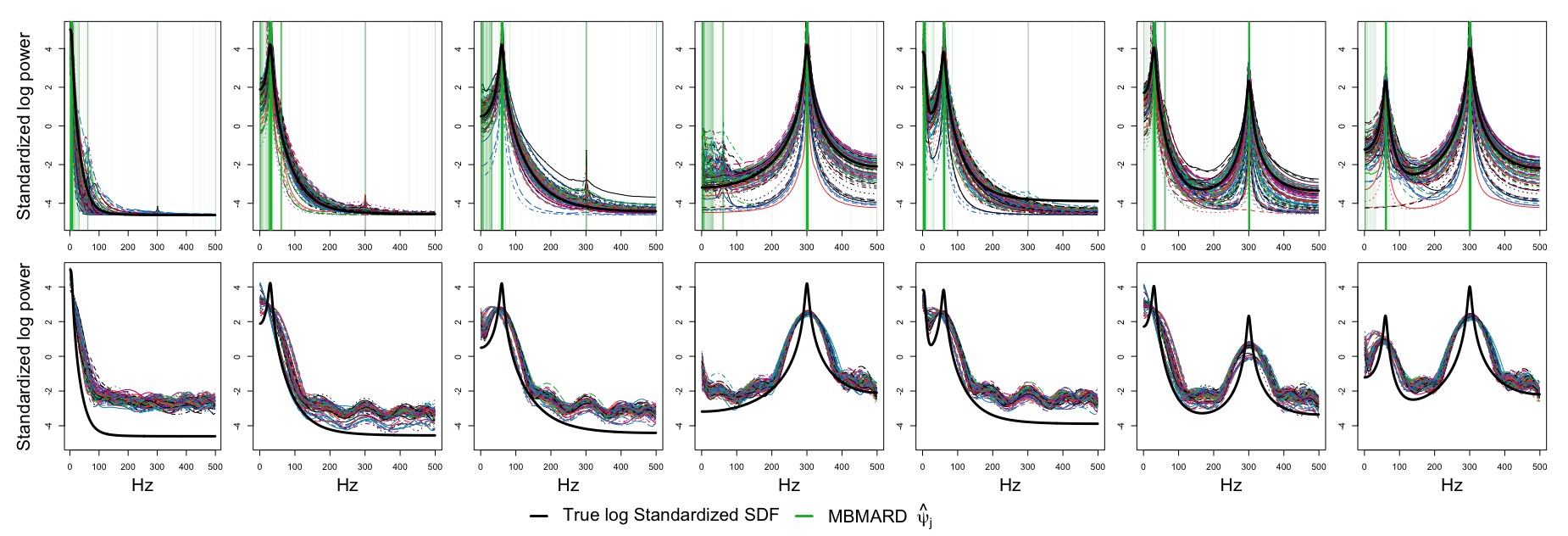}
    \caption{ A mixture of univariate AR(2) processes scenario. Top) 100 estimated curves from the MBMARD method shown in different colors. The green vertical lines correspond to the estimated location parameters with width set as the estimated mixture weights. (Bottom) 100 estimated curves from the Berstein polynomials method. The thick black curve is the logarithm of the true spectral density of each process $X_i(t)$.    }
    \label{fig:AR2sim}
\end{figure}

Since the underlying process is a mixture of AR(2), the first scenario results shown in Figure \ref{fig:AR2sim} demonstrate that MBMARD retrieves the prominent peak's location and shape. The effect of the additional white noise is that MBMARD considers a series of peaks over high frequencies in the final estimator. However, to dismiss the extra high-frequency components, the mixture weights were used to filter only waves with considerable impact on the overall process variance. 

As acknowledged by \cite{MEIER2020104560}, the Bernstein polynomial can produce oversmoothed estimates, which is noted on the broad estimation of the true peaks. The Bernstein polynomial method estimates the white noise effect by smooth curves across the high frequencies. Both methods made a fair estimation of the cross-spectrum of the first scenario (not shown), which is sparse due to the definition of the mixture. 

The misspecification scenario challenges both methods to estimate the spectral matrix and find the main oscillatory components of the processes.  The Bernstein polynomial estimation follows the overall shape of the true spectrum with a series of smooth curves across all the domains. In contrast, the MBMARD estimator shows how the shape of the AR(2) kernel is suitable for autoregressive processes as the AR(12) and AR(1) components. 

As expected, the AR(1) components are identified as latent components present over the signals three to seven. The MA(4) signal points out a limitation since its spectrum has small curves on both sides of the frequency domain that can be identified as components; however, these curves are disregarded, and the AR(1) identification is prioritized. The purpose of the AR(1) components on the mixture is to show that when two peaks share the same location, the latent waves with higher contribution to the total variance, in this case, both AR(1), are identified.
 The proposed MBMARD method decomposes each spectral matrix on a few prominent peaks associated with the single peak on the individual spectral densities. This decomposition for high-order autoregressive processes leads to a good fit, while for moving average processes, the fit is limited due to the autoregressive kernel shape. When the target is to estimate processes with complex associations as in the vector autoregressive scenarios, MBMARD finds the common peaks across the series. However, since the MBMARD retrieves the main latent processes with high spectrum power tends to miss smaller peaks overlapping high power peaks. In theory, when the time series is very long, then there is improved resolution in frequency and thus MBMARD can separate the two peaks in the frequency axis.
 
 
 The Bernstein polynomial is a smoothing technique of the Fourier values. In some situations, the information of small peaks is retrieved, while the peaks are over smoothed in other cases. An important feature required in brain signal analysis is carefully retrieving the frequencies at which only prominent peaks appear since those peaks will be associated with cognitive processes. Then, cases where the method returns wiggle estimates can lead to false-positive detection of peak activity,  for instance, the wiggle estimates of series with AR(1) components. 
\begin{figure} [H]
    \includegraphics[width=16cm, height=6cm]{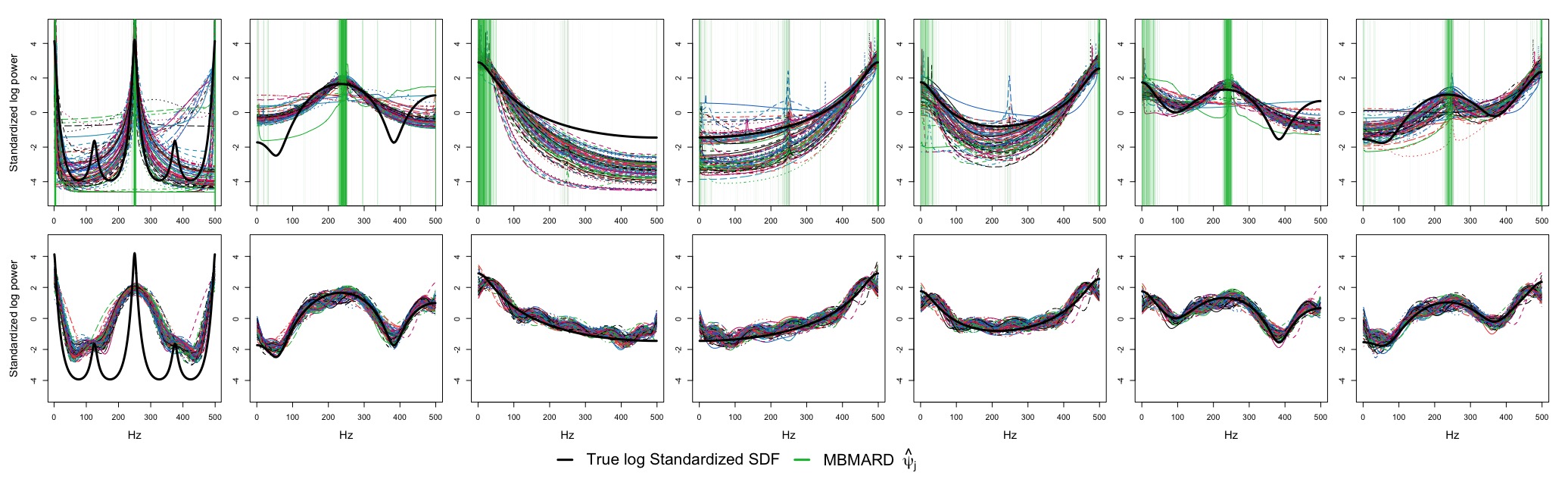}
    \caption{ AR(12), AR(1), MA(4) mixture (misspecification) scenario. Top) 100 estimated curves from the MBMARD method shown in different colors. The green vertical lines correspond to the estimated location parameters with width set as the estimated mixture weights. (Bottom) 100 estimated curves from the Berstein polynomials method. The thick black curve is the logarithm of the true spectral density of each process $X_i(t)$.  }
    \label{fig:multimixsim}
\end{figure}

\begin{figure} [H]
 \centering
    \includegraphics[width=10cm]{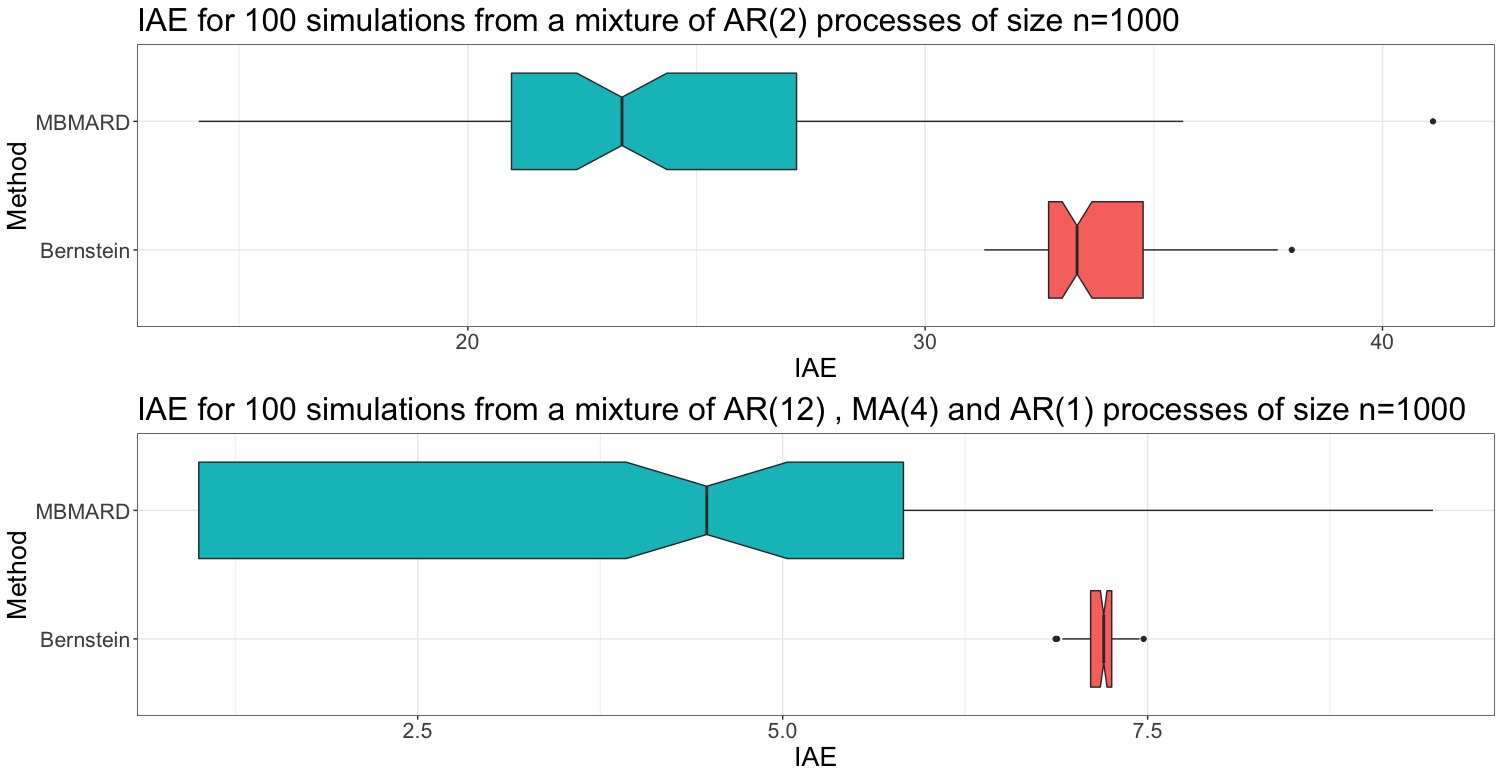}
    \caption{ Boxplots of the integrated absolute error (IAE) of 100 simulations for two settings: (Top) 7 nodes generated from a mixture of AR(2) processes; (bottom) 7 nodes generated from a mixture of an AR(12), MA(4) and two AR(1) processes. The error is displayed for MBMARD (Green) and the Bernstein polynomials method (Orange) }
    \label{fig:IAE}
\end{figure}

The discrepancy between the estimated and true spectral matrices is computed through the integrated absolute error (IAE) which we define as follows. Let $\hat{S}_{XX}^{ij}(\omega)$ and $S_{XX}^{ij}(\omega)$ be the elements of the estimated and true spectral matrices for the observed a multivariate time series $\{X_i(t)\}_{i=1}^n, t=1,\ldots,T$ respectively. Due to $\hat{S}$ and $S$ are hermitian, it is sufficient to compute the error for the lower triangular part of the matrices, then the integrated absolute error is  $\sum_{i \leq j} \int_0^{.5} |\hat{S}_{XX}^{ij}(\omega)-S_{XX}^{ij}(\omega)| d\omega $. The MBMARD estimations outperform the estimations based on Bernstein polynomials as shown in Figure \ref{fig:IAE} boxplots of the IAE for the 100 simulated multivariate signals.

The MBMARD algorithm was run in a HPC cluster formed by Intel Cascade Lakes (40 cores), Skylakes (40 cores), AMD Rome (108 cores) assigned through the workload manager Slurm. The computational performance was, on average, 2 hours and a half for the misspecification scenario and 3 hours and a half for the AR(2) mixture scenario. The Bernstein polynomial method, on average, takes 6 hours to finish 50000 iterations for both scenarios. The difference in computational efficiency is due mainly to the order of updates. MBMARD maintains a small number of kernels by detecting only prominent peaks. At the same time, the Bernstein polynomial relies on increasing the number of kernels to achieve better accuracy in estimating the spectral curves.  MBMARD and the Bernstein polynomial method take around 1 hour and a half to run 50000 iterations in both vector autoregressive scenarios. 

\section{Comparative analysis from EEG of an Alcoholic and a Non-alcoholic subjects}

Memory studies are essential to discover the factors affecting memory performance. For instance, there is evidence of long terms affectations of alcoholism on retrieving visual information (see \cite{zhanga_is_1997, zhang_electrophysiological_1997}  ). However, other studies found alcohol consumption effects on memory seem negligible on retrieving information about words (\cite{zhang_chronic_1997}).

According to \cite{bjork_chapter_1996}, the brain retrieves better the meaning associated to words than trying to remember their superficial features like shape. Another theory on human memory says that neuronal processes driven by consciously encoding memories, called explicit memory, are affected unconsciously by past experiences or implicit memory(see \cite{schacter_implicit_1993}).
Other studies point to evidence that different populations have different signals behavior when performing implicit or explicit memory tasks \citep{bjork_chapter_1996}. Neuroscientists have reached these conclusions by obtaining electroencephalogram (EEG) records as a non-invasive technique to measures electrical signals across the scalp.

The EEG recordings of two subjects, one alcoholic and one non-alcoholic, are analyzed via MBMARD to identify subject-specific dependency within and across brain regions. The MBMARD's properties to decompose multivariate brain signals in its fundamental latent waves allows the comparison of both subjects. This approach can potentially be extended to infer and compare recording from populations using hierarchical modeling, however, that extension goes beyond the goal of this paper that introduces and demonstrate the advantages of this multivariate methodology.

The experiment consists of presenting to each subject images for 300 ms over three different paradigms. In the first task, a single image is presented to the subject during the 3 seconds trial (Single). In the second task is shown twice the same image (Matching), while in the third task, each subject is exposed to two different images (No Match). The data were recorded at a sample rate of 256 Hz for 1 second for ten trials for each task. See \cite{zhang_event_1995} for a detailed and technical description of the data acquisition. 

The MBMARD method was applied over 61 EEG channels for ten trials running 50,000 iterations for three chains leaving a burn-in period of 30,000 iterations. The estimated location parameters $\psi_j$ were grouped for all the trials in three of the classic prespecified frequency bands, delta (0-4 Hertz), Theta (4-8 Hertz), Alpha (8-12 Hertz), Beta (12-30 Hertz), Gamma (30-60 Hertz). 

\begin{figure}[H]
    \centering
    \includegraphics[width=12cm]{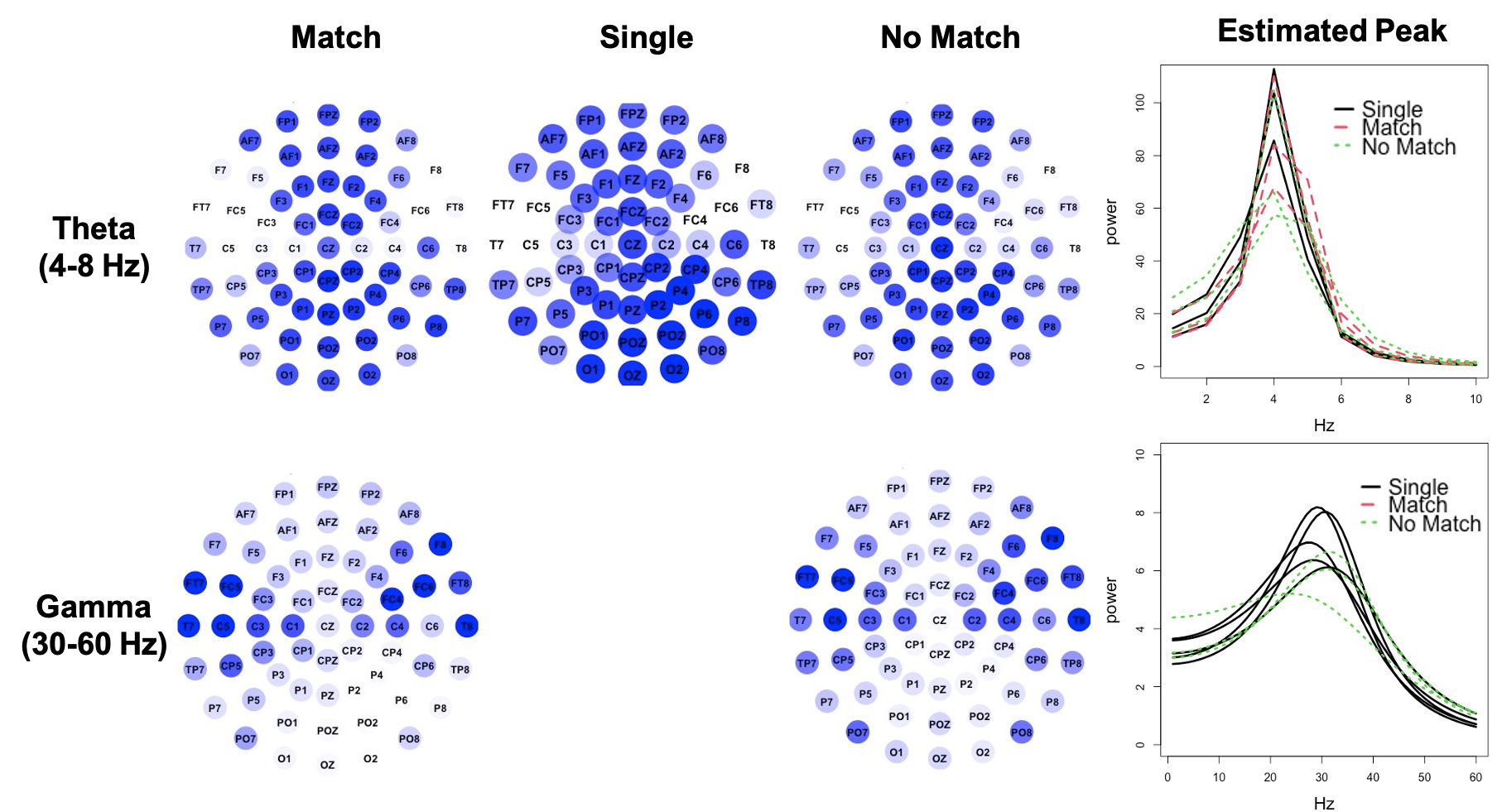}
    \caption{The circular plots show the identified oscillatory components in an alcoholic subject during an image recognition experiment. The estimated mixture weights determine the color opacity of the nodes. The Theta component is active in the frontal and occipital regions oscillating around 4.3 HZ, while the Gamma component (30-60 HZ) is active in the temporal regions centered at 32 Hz}
    \label{fig:alchsum}
\end{figure}

The alcoholic subject shows two principal spectral peaks. The first peak is located in the theta band. Figure~\ref{fig:alchsum} shows that the theta component is a sharp peak centered around 4-5 Hertz. The Theta peak appears over the frontal and occipital zones in some trials for all the experimental conditions. The Second peak lies on the Gamma band, with higher weights in both temporal zones, but is only found in the Single and No match trials. Contrary to the Theta peak, the Gamma peak is a broad peak with more variability in its centers. Table \ref{table1} contains the estimated average parameters of the peaks extracted from the alcoholic subject EEG data for each task.

\begin{table}
	\centering
	\begin{tabular}{c|c|c|c|c|c}
		\hline
		Band & Task & $\bar{\psi}$ (Hz)  & $\bar{L}$ & $\bar{\lambda}$ & \# Trials  \\
		\hline
		\multirow{3}[2]{*}{Theta} & Single & 4.25 & 0.02 & .65 & 3  \\
		& Match & 4.44 & 0.02 & .68 & 3 \\
		& No Match & 4.35 & 0.03 & .67 & 3 \\
		\hline
		\multirow{2}[2]{*}{Gamma} & Single & 31.7 & 0.32 & .47 & 5  \\
		& No Match & 32.8 & 0.41 & .48 & 3 \\
		\hline
	\end{tabular}%
		\caption{Summary statistics of the peaks found for the Alcoholic subject. The average was computed for the location $\bar{\psi}$, bandwidth $\bar{L}$, and mixture weight $\bar{\lambda}$  The last columns contains the number of trials where the peak was found.\label{table1}}
\end{table}%
The Non-alcoholic subject shows a low-frequency component found in the Alpha band for the occipital brain region. This peak has high variability concerning its central location, bandwidth, and weight. Despite being found in several trials, the Alpha component indicates a non-stable behavior of the Alpha oscillations.

The Gamma band contains the second identified peak. However, the Gamma peak is such broad that the latent process estimated includes several oscillations at different frequencies. It is noted that only the peak during a No Match trial is sharp enough to be associated with a quasiperiodic process. The summary of the parameters of the peaks for the Non-alcoholic subject is displayed in Table \ref{table2}.

The low-frequency components are consistent with the neuroscientific literature since according to \cite{dicarlo_how_2013} the ventral stream, which is located in the occipital and temporal zone, is where our brain processes object recognition. 

Another analysis of the alcoholics and healthy subjects data set has centered on two goals. The first approach is to develop methods that use the EEG recordings or their features as inputs of classification algorithms to predict if a subject is an alcoholic or healthy subject. However, these methods do not delve into the features that help to differentiate between subjects and focus only on accuracy performance (See \cite{palaniappan_vep_2002}, \cite{kok_meng_ong_selection_2005}, \cite{yazdani_classification_2007}, \cite{guntaka_eeg_based_2013}).

\begin{figure}[H]
    \centering
    \includegraphics[width=12cm]{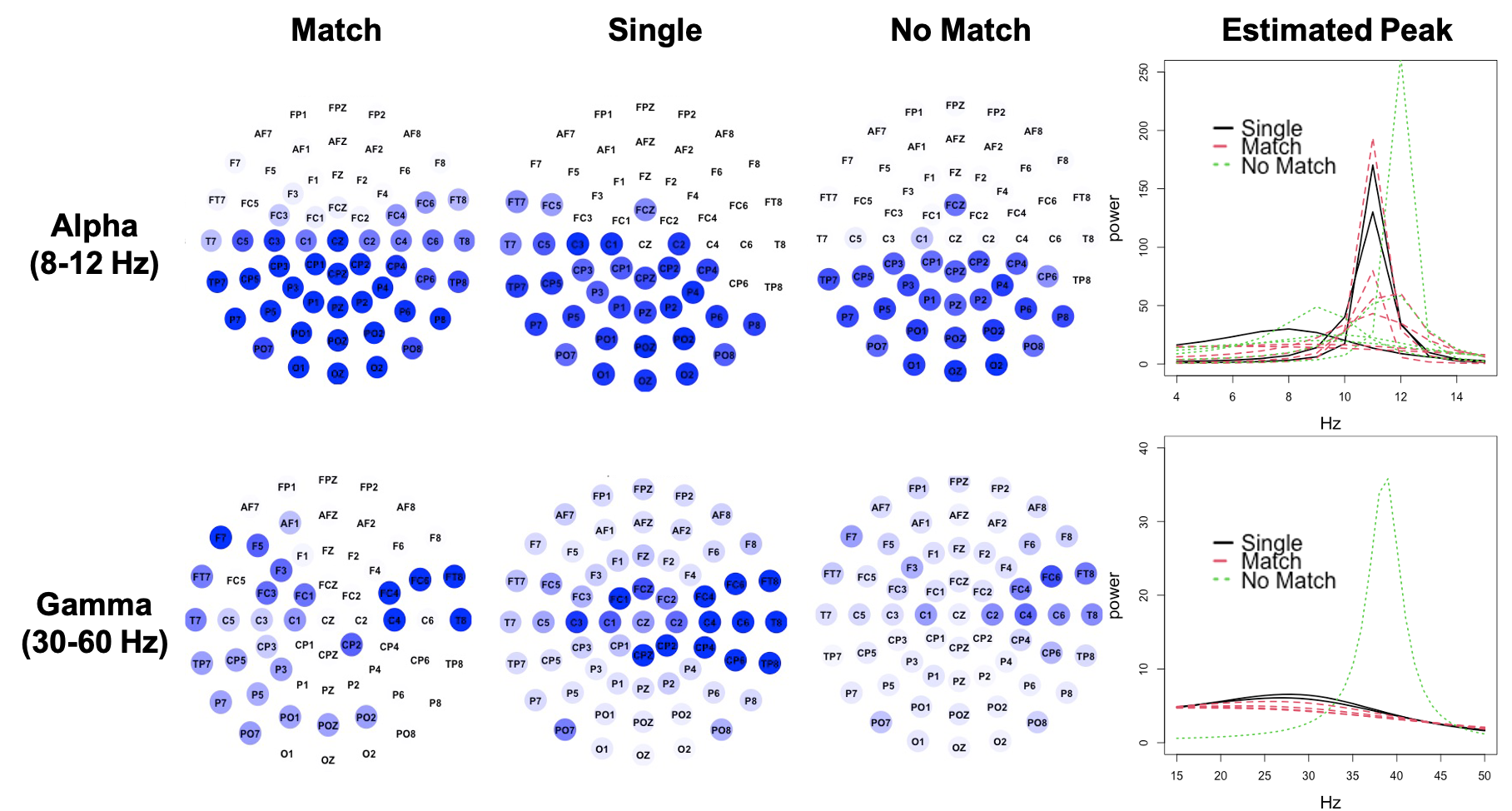}
    \caption{The circular plots show the identified oscillatory components in an alcoholic subject during an image recognition experiment. The estimated mixture weights determine the color opacity of the nodes. The Alpha component is active in the occipital region oscillating around 10-11 HZ, while the Gamma component (30-60 HZ) is active in the temporal regions centered at 30.6 Hz}
    \label{fig:fitmatching}
\end{figure}
\begin{table}
	\centering
	\begin{tabular}{c|c|c|c|c|c}
		\hline
		Band & Task & $\bar{\psi}$ (Hz)  & $\bar{L}$ & $\bar{\lambda}$ & \# Trials  \\
		\hline
		\multirow{3}[2]{*}{Alpha} & Single & 10.2 & 0.03 & .34 & 3  \\
		& Match & 11 & 0.08 & .64 & 6 \\
		& No Match & 10.7 & 0.05 & .61 & 5 \\
		\hline
		\multirow{3}[2]{*}{Gamma} & Single & 30.7 & 0.38 & .47 & 2  \\
		& Match & 30.6 & 0.6 & .28 & 4 \\
		& No Match & 38.7 & 0.05 & .15 & 1 \\
		\hline
	\end{tabular}%
		\caption{Summary statistics of the peaks found for the Alcoholic subject. The average was computed for the location $\bar{\psi}$, bandwidth $\bar{L}$, and mixture weight $\bar{\lambda}$  The last columns contains the number of trials where the peak was found.\label{table2}}
\end{table}%

The second approach decomposes each channel signal according to frequency bands, analyzing each band's power and making a comparative analysis. For instance, \cite{sun_eeg_2006} uses PCA for dimension reduction and then applies a  wavelet decomposition into five frequency bands, finding an increased theta and delta power and a lower power in the alpha band for alcoholics subjects. However, they used different frequency bands, making it difficult to compare with analysis and reinforce the idea or use frequency-specific methods to identify the signals waveforms. 

\cite{tcheslavski_alcoholism_related_2012} comprehensively analyzed frequency bands difference in spectral power, coherence, and phase synchrony frequency bands between alcoholics and non-alcoholics, evaluating the significance of the differences using Kruskall-Wallis tests. The authors found that alcoholic's lower rhythms have lower spectral power and coherence. Delta and Theta rhythms show higher differences in the frontal region, while bands from 8-20 Hz highlight differences in the occipital region. 

Our analysis is consistent with previous reports by finding that low-frequency waveforms contribute the highest to the spectral matrix in the frontal and occipital zones. Besides, an additional component was found in the gamma band on the temporal zones. Individual runs of our algorithm to the rest of the subjects in the data set (not shown here) are consistent with our previous findings. However, variability exists across subjects in the peak locations, bandwidths, and weights that should be treated in a hierarchical model. 

In brief, the contributions of MBMARD to the alcoholic and healthy subject analysis are 1) the joint modeling of the EEG nodes finding shared information and dependence among the signals, 2) The identification of frequency-specific waveforms in the signals, zooming the analysis into particular frequencies, and making the segmentation by frequency bands optional for comparison with other analysis.

\section{Conclusion}

The Multivariate Bayesian Mixture Autoregressive Decomposition (MBMARD) is a novel Bayesian non-parametric method that employs the spectral densities of an AR(2) as an unimodal kernel that decompose a multivariate time series into latent stochastic processes.

The parametric kernel allows the straightforward derivation of classical dependence measures as cross-correlation and PDC between the signals in terms of the model parameters, providing a tool to display the association between the channels from the shared latent waveforms. 

The Simulations study shows that MBMARD generates a smooth estimator of the spectral density matrix finding the most prominent peak activity of the signals. The computational performance of MBMARD has shown to be more efficient due to maintaining a lower number of components.

The analysis of two subject's 61 channels EEG dataset via MBMARD highlighted two components in the signals low-frequency and high-frequency oscillations, activated in different zones of the brain. The MBMARD analysis isolated these component's main characteristics, finding differences in frequency location, shape, variability, and contribution to the EEG variance. MBMARD provides interpretable properties to analyze brain signals under different contexts. In addition, the straightforward understanding of the time series decomposition makes the proposed method applicable to different contexts in time series analysis. 

\bibliographystyle{chicago}
\typeout{}
\bibliography{BibliographyMMMC}

\section*{Appendix}

The Partial directed coherence of the MBMARD model can be computed explicitly, since the MBMARD model has the same structure as a vector autoregressive model by assuming the signals are a convex combination of the latent oscillations. 

The  PDC matrix based on the notation of \citeauthor{baccala_partial_2001} is derived in the following way. First, we define the matrix $A(\omega)$ as the sum of Fourier transform of the coefficient matrices $A_1$ and $A_2$ corresponding to the first and second-order lag, respectively, giving the following block matrix to then compute $I-A(\omega)$

\[
A(\omega) =
\begin{pmatrix}
A_1e^{-i2\pi\omega}+A_2e^{-i4\pi\omega} &  0_{K \times n} \\
\Lambda (A_1e^{-i2\pi\omega}+A_2e^{-i4\pi\omega})  &  0_{n \times n}
\end{pmatrix}
\]

\[
\bar{A}(\omega)=\{\bar{A}_{ij}(\omega) \}=I-A(\omega) =
\begin{pmatrix}
I_{K \times K}- A_1e^{-i2\pi\omega}-A_2e^{-i4\pi\omega} &  0_{K \times n} \\
-\Lambda (A_1e^{-i2\pi\omega}+A_2e^{-i4\pi\omega})  &  I_{n \times n}
\end{pmatrix}
\]

The bottom left block from the matrix $I-A(\omega)$ is the influence of the latent oscillations $Z(t)$ to the signals $X(t)$ expressed as follows,

\[
-\Lambda (A_1e^{-i2\pi\omega}+A_2e^{-i4\pi\omega}) = -
\begin{pmatrix}
\lambda_{11}(\phi_{11}e^{-i2\pi\omega}+ \phi_{12}e^{-i4\pi\omega}) & \dots & \lambda_{1K}(\phi_{K1}e^{-i2\pi\omega}+ \phi_{K2}e^{-i4\pi\omega}) \\
\vdots & \dots &\vdots \\
\lambda_{n1}(\phi_{11}e^{-i2\pi\omega}+ \phi_{12}e^{-i4\pi\omega}) & \dots & \lambda_{nK}(\phi_{K1}e^{-i2\pi\omega}+ \phi_{K2}e^{-i4\pi\omega})  
\end{pmatrix}
\]

All the columns have the same factor due to $A_1$ and $A_2$ are diagonal matrices. Denote the common factor per column as $\Phi_j(\omega)=\phi_{j1}e^{-i2\pi\omega}+ \phi_{j2}e^{-i4\pi\omega}$ for $j=1,\ldots,K$. To standarize the PDC values we denote as $a_j(\omega)$ the $j$-th column of the matrix $\bar{A}(\omega)$  then the PDC from the $Z_j(t)$ to $X_i(t)$ ($Z_j \rightarrow X_i$) is computed as:

\[
|\pi{ij}(\omega)|= \frac{|\bar{A}_{ij}(\omega)| }{\sqrt{a_j^H(\omega)a_j(\omega) } }
\]

where $H$ indicate the transpose conjugate. in terms of the parameters the denominator is derived as

\[
a_j^H(\omega)a_j(\omega)= |\Phi_j(\omega) |^2 \sum_{i=1}^n \lambda^2_{ij} +|1-\Phi_j(\omega)|^2 
\]

where the right term in the sum comes from the diagonal matrix in the upper left block from $\bar{A}(\omega)$. The numerator of the PDC is $|\bar{A}_{ij}(\omega)|=\lambda_{ij} |\Phi_j(\omega)|$ therefore, we can simplify the case when is of interest the PDC from the latent oscillation $Z_j$ to the signal $X_i$ in terms of the model parameters through
\[
|\pi_{ij}(\omega)|= \frac{\lambda_{ij} |\Phi_j(\omega)| }{\sqrt{|\Phi_j(\omega) |^2 \sum_{i=1}^n \lambda^2_{ij} +|1-\Phi_j(\omega)|^2  } }
\]

The other case of interest is the PDC between the signals; however, derived from the identifiability conditions of the model the bottom right side of the block matrix $\bar{A}(\omega)$ is the identity that leads to the PDC from signal $X_m$ to $X_l$ ($X_m \rightarrow X_l$) be defined as
\[
|\pi_{lm}(\omega)| = 
     \begin{cases}
      \text{1,} &\quad\text{if} \ m=l \\
      \text{0,} &\quad\text{if} \ m \ne l  \\
     \end{cases}
\]

Therefore, all the associations between the signals are given through their shared latent waves.

\end{document}